\documentclass[preprint2]{aastex}

\usepackage{rotating}

\newcommand{\hd}{HD\,148937}

\newcommand{\myemail}{naze@astro.ulg.ac.be}
\newcommand{\kms}{km\,s$^{-1}$}
\newcommand\nspec[4]{\mbox{$#1\,^{#2}{#3}_{#4}$}}

\begin{document}

\title{High-resolution X-ray spectroscopy of the magnetic Of?p star \hd}

\author{Ya\"el Naz\'e\altaffilmark{1}}
\affil{GAPHE, D\'epartement AGO, Universit\'e de Li\`ege, All\'ee du 6 Ao\^ut 17, Bat. B5C, B4000-Li\`ege, Belgium}
\email{\myemail}

\author{Svetozar A. Zhekov}
\affil{JILA, University of Colorado, Boulder, CO 80309-0440, USA ; Space and Solar-Terrestrial Research Institute, Sofia, Bulgaria}
%\email{zhekovs@colorado.edu}

\and

\author{Nolan R. Walborn}
\affil{Space Telescope Science Institute, 3700 Martin Drive, Baltimore, MD 21218, USA}

\altaffiltext{1}{FRS-FNRS Research Associate}

\begin{abstract}
High-resolution data of the peculiar magnetic massive star \hd\ were obtained with Chandra-HETGS, and are presented here in combination with a re-analysis of the older XMM-RGS data. The lines of the high-Z elements (Mg, Si, S) were found to be unshifted and relatively narrow (FWHM of about 800\,\kms), i.e. narrower than the O line recorded by RGS, which possibly indicates that the hot plasma is multi-thermal and has several origins. These data further indicate a main plasma temperature of about 0.6\,keV and a formation of the X-ray emission at about one stellar radius above the photosphere. From the spectral fits and the H-to-He line ratios, the presence of very hot plasma is however confirmed, though with a smaller relative strength than for the prototype magnetic oblique rotator $\theta^1$\,Ori\,C. Both stars thus share many similarities, but \hd\ appears less extreme than $\theta^1$\,Ori\,C despite having also a large magnetic confinement parameter.
\end{abstract}


\section{Introduction}
The peculiar class of Of?p stars has received much attention in recent years. These objects were first distinguished from normal Of supergiants some 40 years ago \citep{wal72} because of the unusual presence of C\,{\sc iii}\,$\lambda$\,4650\AA\ in emission with a strength comparable to the neighbouring N\,{\sc iii} lines. In the last decade, additional peculiar properties were discovered such as recurrent spectral variations (in the Balmer, He\,{\sc i}, C\,{\sc iii}, Si\,{\sc iii} lines), large X-ray overluminosities, and the presence of strong magnetic fields, leading to a refinement in the definition of the class \citep[see][and references therein]{naz08b,wal10}. In total, five Of?p stars are now identified in the Galaxy.

\hd\ was one of the first two Of?p identified \citep{wal72}, but it was only studied in detail recently. Contrary to other members of the Of?p category, this star does not present large photometric changes or impressive variations of the optical spectra occurring on very long periods (e.g. 538d for HD\,191612, $\sim$55yrs for HD\,108). Instead, smaller-amplitude variations of the optical spectrum occurring on shorter timescales (7\,d) have been detected \citep{naz08,naz10}. The observed magnetic field of \hd, first reported by \citet{hub08} and recently confirmed by \citet{wad11}, also seems to present no or only very small variations over the 7d-period. Such a lack of large variations, in contrast with other objects of the same category, does not necessarily rule out the Magnetic Oblique Rotator (MOR) model that is proposed to explain the Of?p peculiarities. Small variations could indeed be attributed to a pole-on geometry, as seen from Earth \citep{naz10,wad11}, though such a geometry would be at odds with the appearance of the surrounding ejected nebula.

In the X-ray domain, the high-energy emission of \hd\ was first detected with Einstein, and then re-observed by ROSAT and XMM-Newton \citep[and references therein]{naz08}. As for the two other famous members of the Of?p category, HD\,108 and HD\,191612, the spectrum of \hd\ appeared at the same time overluminous ($\log (L_X/l_{\rm BOL})\sim-6$ rather than --7, \citealt{naz08,naz09,naz11}) and rather soft (dominant plasma temperature at 0.2 or 0.6 keV). These properties are at odds with what is expected from both magnetic objects such as $\theta^1$\,Ori\,C \citep[and references therein]{naz10} and typical O-stars. To go further into the analysis and fully characterize the physical properties of \hd, high-resolution data of high quality are needed. However, the high-resolution data from the RGS instrument aboard XMM-Newton were very noisy, though they seemed to indicate the presence of broad X-ray lines (FWHM$\sim$2000\kms), in agreement with prediction from the `standard' wind-shock model of O-stars \citep{naz08}. This is indeed puzzling since \hd\ and the other Of?p stars have magnetic confinement parameters as large as $\theta^1$\,Ori\,C \citep{naz08b} and one would thus expect rather narrow X-ray lines. New data were thus essential.

This paper reports on the analysis of Chandra high-resolution data and for completeness we re-visit the XMM-Newton RGS data. The observations are presented in Sect. 2, the properties of the X-ray lines and overall high-energy emission are analyzed in Sects. 3 and 4, respectively, while Sect. 5 summarizes the results and concludes.

\section{Observations and data reduction}
\subsection{Chandra}
\hd\ was observed by Chandra on June 5 2010 using the HETG+ACIS-S (ObsID=10982, PI Y. Naz\'e). The available pipeline data (level 2) were not re-processed, as they had been reduced with a recent reduction software. They were further reduced with CIAO 4.2 and CALDB 4.3.1. Following the recommendations of the Chandra team\footnote{http://cxc.harvard.edu/ciao/threads/gspec.html}, we first extracted individual first-order MEG and HEG spectra for the source and background using {\it dmtype2split} and {\it tg\_bkg}, then calculated response matrices for both first-orders using {\it mkgrmf} and {\it fullgarf}. The spectra and arf files from orders --1 and +1 were added for each instrument using {\it add\_grating\_orders}, and the rmf file for the +1 order was used as general response matrix for each instrument (since the differences between the two rmf are very small). For the 0th order data, we further defined a source region of 15 pixels radius and a surrounding annular background region of radii 15 and 45 pixels. We then used the task {\it specextract} to get the 0th order spectra and its corresponding response matrices. 

The total effective exposure was 99ks, and the observed count rates in the 0.5--2.5\,keV energy band are $1.97 \pm 0.17 \times10^{-2}$\,cts\,s$^{-1}$, $5.31 \pm 0.15 \times10^{-2}$\,cts\,s$^{-1}$, and $4.41 \pm 0.07 \times10^{-2}$\,cts\,s$^{-1}$ for HEG, MEG, and 0th order, respectively. In the 0.5--10.\,keV range, the count rates (resp. net number of counts) are $2.43 \pm 0.18 \times10^{-2}$\,cts\,s$^{-1}$ (resp. $\sim$2400), $5.88 \pm 0.16 \times10^{-2}$\,cts\,s$^{-1}$ (resp. $\sim$5800), and $5.90 \pm 0.08 \times10^{-2}$\,cts\,s$^{-1}$ (resp. $\sim$5800) for HEG, MEG, and 0th order, respectively. It may be noted that count rates in the  0.5--10.\,keV range are significantly higher than those in the 0.5--2.5\,keV energy band, indicating the presence of some hard energy flux. This could be due either to a high absorption or the presence of very hot plasma, which is actually the case here (see below).

For the line analysis, HEG and MEG spectra were re-binned by a factor of 4 and 2, respectively, to improve the signal-to-noise without degrading the resolution. For the global fitting, the HEG, MEG and 0th order spectra were grouped to get a minimum of 20 counts per bin. Following \citet{zhp07}, the MEG and HEG spectra near the intercombination and forbidden lines in the helium-like triplets were re-binned so that these two lines fall into one large bin. This avoids fitting problems as Xspec thermal models (e.g. apec) do not take into account the potential $f/i$ variations if a strong UV emission is present. For both binnings, all available data (HAG+MEG for line fitting and HEG+MEG+0th order for global fitting) were fitted simultaneously, using a $\chi^2$ statistics and background subtraction.

\subsection{XMM-Newton}
XMM-Newton observed \hd\ on Feb. 25 2001. Three archival datasets with RGS data exist, with ObsID 0022140101, 0022140501, 0022140601. Compared to \citet{naz08}, the XMM reduction software has significantly improved, especially for the grating data. The full observation was therefore downloaded and reduced once again with SAS v10.0.0, following the recommendation of the XMM team\footnote {http://xmm.esac.esa.int/sas/current/howtousesas.shtml}. The data from each instrument and each order were combined using the task {\it rgscombine}. A high background affects all three exposures, but without a clear flare: the whole exposure (38ks for RGS1-order1 and 36ks for RGS2-order1) was thus kept. The overall count rates in the 0.5--2.5\,keV energy band are $3.71 \pm 0.25 \times10^{-2}$\,cts\,s$^{-1}$ and $6.69 \pm 0.25 \times10^{-2}$\,cts\,s$^{-1}$ for the first orders of RGS1 and RGS2, respectively. This corresponds to net number of counts of about 1400 and 2400 for the first orders of RGS1 and RGS2, respectively. A grouping of the channels by a factor of 3 was used, to improve the signal-to-noise without degrading the resolution. Note that (1) only first-order data were considered since the second-order data are very noisy; (2) only RGS2 data are available at the wavelength of the neon lines because of a CCD failure in RGS1.

\section{Line properties}
\begin{figure}
\includegraphics[width=8cm]{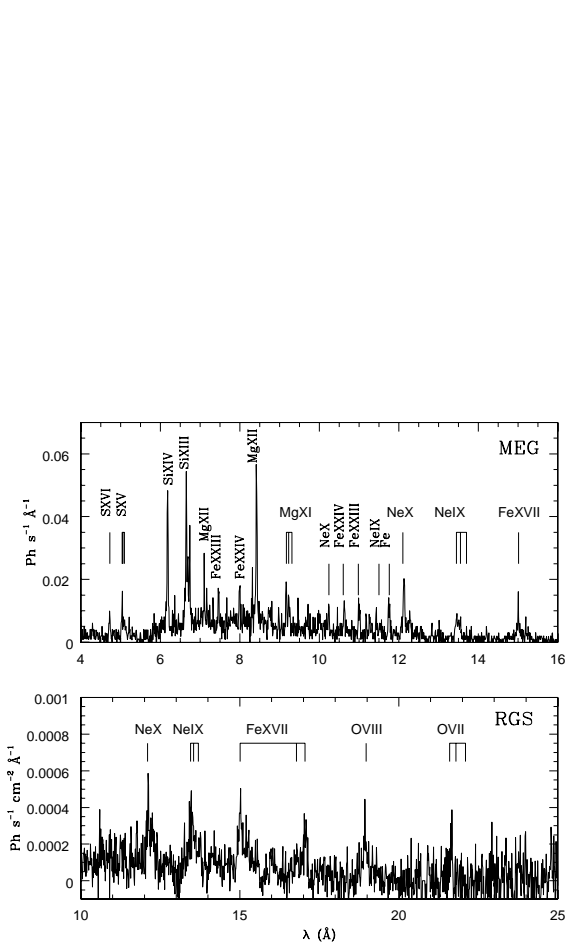}
\caption{The MEG and RGS spectra with the strongest lines identified. Note that (1) the wavelength range is not identical in both panels because of the different sensitivities of the considered instruments and (2) the spectra have been background subtracted. \label{spec}}
\end{figure}

Both RGS and MEG spectra are shown in Fig. \ref{spec}. The strongest lines are Ne\,{\sc x}\,$\lambda$12\AA\ and O\,{\sc viii}\,$\lambda$19\AA\ for the RGS and S\,{\sc xv}\,$\lambda$5\AA, Si\,{\sc xiv}\,$\lambda$6.2\AA, Si\,{\sc xiii}\,$\lambda$6.7\AA, Mg\,{\sc xii}\,$\lambda$8.4\AA, Mg\,{\sc xi}\,$\lambda$9.2\AA, and Ne\,{\sc x}\,$\lambda$12\AA\ for the MEG. It may be noted that lines of highly-ionized iron are detected, which is not common for `normal' O stars.

\begin{figure}
\includegraphics[width=8cm]{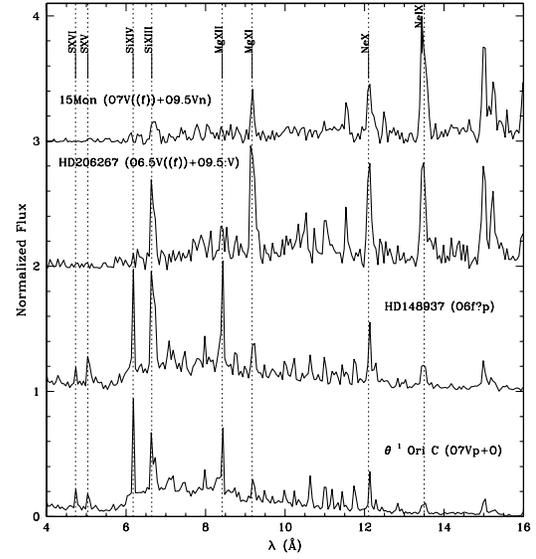}
\caption{MEG spectra of a selection of OB stars (spectral channels were binned by a factor of ten for all stars, as in \citealt{wal09}). Actual fluxes have been normalized so that the strongest lines peak to 1, and the most important H- and He-like pairs are labelled. \label{global}}
\end{figure}

In Fig. \ref{global}, we also compare the spectrum of \hd\ to those of the magnetic object $\theta^1$\,Ori\,C (ObsID 3,4), and two `normal' O stars, 15\,Mon (ObsID 5401,6247) and HD\,206267 (ObsID 1888,1889), from the sequence of \citet{wal09}. Of these four stars, 15\,Mon suffers from the smallest interstellar absorption (E(B-V)=0.03), while $\theta^1$\,Ori\,C is more extincted with E(B-V)=0.29 and the last two objects display the highest extinctions, with E(B-V)$\sim$0.5. This implies that differences between HD\,206267 and \hd\ cannot be due to different interstellar absorptions, and that the hard character of $\theta^1$\,Ori\,C is not due to a high extinction.

From Fig. \ref{global}, several features are obvious. The Ne\,{\sc x} line largely dominates over the Ne\,{\sc ix} line in the spectra of $\theta^1$\,Ori\,C and \hd. Though this pair of lines is the most separated one in wavelength, hence potentially the most affected by absorption effects, this difference is not due to interstellar absorption, as \hd\ and HD\,206267 share similar interstellar extinctions. A similar feature is seen for the H-to-He line ratio of magnesium. Regarding silicon, the Si\,{\sc xiv} line is not clearly detected in the ``normal'' O-stars, contrary to Si\,{\sc xiii}, while it is of similar strength to Si\,{\sc xiii} in \hd\ and even dominates over Si\,{\sc xiii} for $\theta^1$\,Ori\,C. Finally, the sulfur lines do not appear in the spectra of ``normal'' O-stars (at the sensitivity limit of the data), but are obvious in $\theta^1$\,Ori\,C and \hd. It should be noted however that S\,{\sc xiv} dominates over S\,{\sc xv} in \hd, whereas the reverse situation is seen in $\theta^1$\,Ori\,C.

The reversal of the S, Si, Mg and Ne H-to-He line ratios between normal stars and magnetic objects is thus clearly seen, the latter having higher ionization. However, it is less extreme for \hd\ than in the case of $\theta^1$\,Ori\,C. \hd\ may nevertheless participate in the dichotomy discussed by \citet{wal09} and \citet{zhp07}.

\subsection {Shifts and widths }
The rest wavelengths of these lines were taken from ATOMDB\footnote{http://www.atomdb.org/}, and Xspec v12.6.0.q was used for the fitting. For He-like triplets (S\,{\sc xv}, Si\,{\sc xiii}, Mg\,{\sc xi}), the shifts and widths of the lines were forced to be equal for the 3 components; whereas for the H-like doublets (Si\,{\sc xiv}, Mg\,{\sc xii}, Ne\,{\sc x}, O\,{\sc viii}), the relative line intensities were in addition tied to their atomic data value at the peak emissivity.

\begin{figure}
\includegraphics[width=8cm]{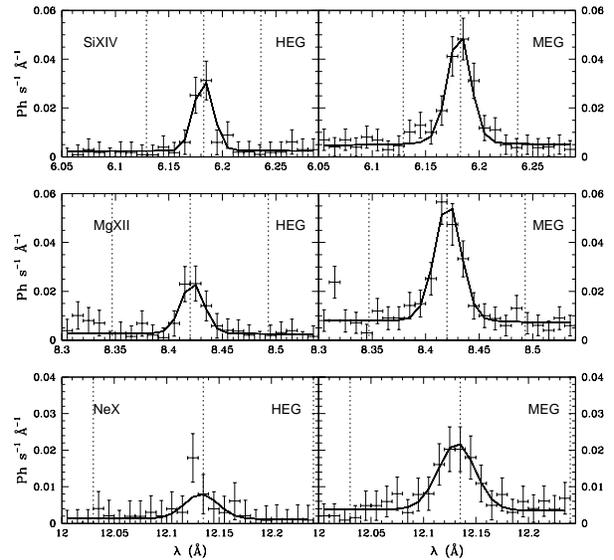}
\caption{The observed H-like line profiles and their best fit models. Dotted lines are drawed at velocities of $-v_{\infty}$, 0 and $+v_{\infty}$. \label{lineH}}
\end{figure}

\begin{figure}
\includegraphics[width=8cm]{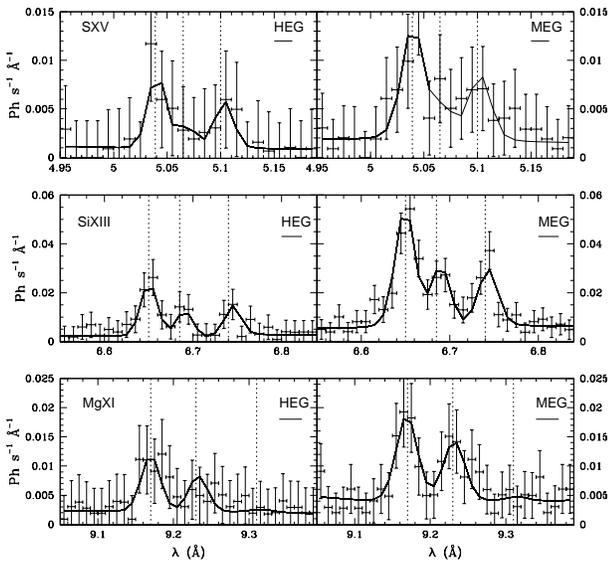}
\caption{The observed He-like line triplets and their best fit models. Dotted vertical lines indicate the rest wavelengths of the $rif$ triplets, and the small horizontal line below the instrument's name has a length of 1000\kms. \label{lineHe}}
\end{figure}

The signal-to-noise of the data prevents us from doing detailed line profile modelling: the lines were thus fitted by simple gaussians. The fitting results are provided in Table \ref{linefit} and shown on Figures \ref{lineH} to \ref{lineHe}. For each line, the rest wavelength/energy, shift, width, and intensities are tabulated, together with their associated $1\sigma$ error. Note that these errors are often asymmetrical: the value shown here always is the largest value. The last column provides the intensities corrected for an interstellar absorption of 4$\times$10$^{21}$\,cm$^{-2}$ \citep{dip94,mormcc}. Note that a fit by a triangular profile was also performed, for consistency with \citet{naz08}: the results are indistinguishable from those reported in Table \ref{linefit}.

The results for the Chandra data are rather stable from line to line: the average shift is $49\pm32$\,\kms, and the average FWHM is $827\pm90$\,\kms\ (note that quoted errors are errors on the mean, not dispersions around the mean). The lines are thus not significantly blue- and only slightly red-shifted, in agreement with the null shift found by \citet{naz08}. This is not surprising since, apart from a few objects \citep{kah01,rau02,coh11}, `normal' O-stars generally show unshifted lines (see \citealt{wal07,gue09} and references therein). The observed widths can also be compared to values reported for massive stars in the literature (see \citealt{gue09} and references therein): the first O-stars observed at high-resolution yielded mixed results, some having broad lines (FWHM of 2000\kms\ for $\zeta$\,Pup and 1700\kms\ for $\zeta$\,Ori), other narrower lines (FWHM of 860\kms\ for $\delta$\,Ori, 800\kms\ for $\zeta$\,Oph); while global studies later showed that FWHM$\lesssim v_{\infty}$ for most stars. Indeed, the average width found here is below the wind terminal velocity (see Table \ref{param}), in line with the ``narrow'' cases mentioned above, but it is not as extreme as the FWHM=600\kms\ reported for most lines of $\theta^1$\,Ori\,C \citep[and erratum in \citealt{gag05b}]{gag05}.

\begin{figure}
\includegraphics[width=8cm, bb= 20 500 590 710, clip]{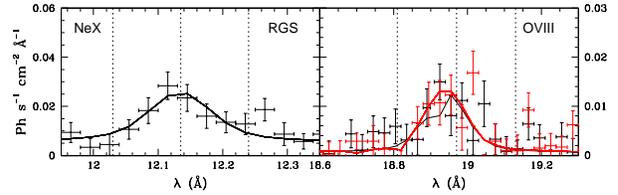}
\caption{The line profiles observed with RGS  and their best fit models. The left panel shows the RGS2 data for Ne\,{\sc x} while the right panel displays both RGS for O\,{\sc viii}, the thin black and thick red line corresponding to RGS 1 and 2, respectively. Dotted lines are drawed at velocities of $-v_{\infty}$, 0 and $+v_{\infty}$. \label{linergs}}
\end{figure}

\begin{figure}
\includegraphics[width=8cm]{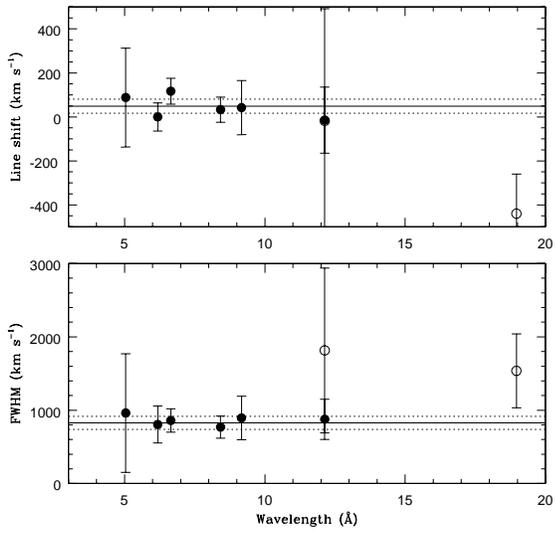}
\caption{The best-fit FWHMs and line shifts. Filled circles refer to Chandra data, open circles to RGS; horizontal lines indicate the average values and their associated $\pm1\sigma$ interval. \label{shiftwidth}}
\end{figure}

However, the line widths measured for \hd\ on Chandra data are much smaller than those reported for the same star by \citet[about 2000\kms]{naz08} using XMM-Newton RGS data. Since the reduction software has significantly changed in the last years, we have re-reduced the RGS data and re-fit the lines (see Sect 2.2, Figs. \ref{linergs} \& \ref{shiftwidth} and Table \ref{linefit}): once again, the oxygen line appears resolved and we found large values for the widths. The differences thus appear real, but two remarks must be made: (1) the signal-to-noise of the XMM data is much lower than that of the Chandra data, which could blur the results (indeed, the error bars are much larger, see  Fig. \ref{shiftwidth}!), and (2) the O\,{\sc viii} line is generally excited by lower-temperature plasma. As the X-ray emission is most probably not isothermal (see next section), the data difference could simply reflect the presence of several plasmas. Indeed, in the case of $\theta^1$\,Ori\,C \citep{gag05,gag05b}, the O\,{\sc viii} width was found to be larger than that of shorter wavelength lines by a factor of 2.5 (which is 40\% higher than what we found for \hd). Therefore, what the Chandra data tell us is that the high-temperature plasma, most probably linked to magnetic confinement close to the star, displays narrow lines, in agreement with theoretical predictions (see e.g. \citealt{udd}). This solves at least one problem in the understanding of the X-ray emission from Of?p stars. 

\subsection{Ratios}
Using the line intensities, we calculated the ${\cal R}=f/i$, ${\cal G}=(f+i)/r$, and H-to-He-like ratios, which we compare to theoretical values for deriving plasma location and temperature. 

We extracted the theoretical emissivity information for the H-like and He-like lines from ATOMDB v2.0.0 \citep{fos11}. Note that ATOMDB calculations are done considering collisional ionization equilibrium and that the used emissivities are not limited to those of the lines themselves as several other lines blend with the doublets and triplets. For our calculations, we have therefore summed the emissivities of all lines found over an interval $\lambda_0\pm0.5\times {\rm resolution}$ around the lines of interest, using resolutions of 1.2$\times10^{-2}$\AA, 2.4$\times10^{-2}$\AA, and 6$\times10^{-2}$\AA\ for HEG, MEG and RGS1-order1, respectively, which correspond to the FWHMs of the instrumental broadening. 

The observed H-to-He-like ratios of Si and Mg favor temperatures $\log(T)$ of 7.08$\pm$0.02 and 6.92$\pm$0.05, respectively. These high temperatures (12 and 8MK) are larger than those of ``normal'' OB stars quoted in \citet[see also the erratum in \citealt{wal08}]{wal07}, but similar to those reported for the colliding wind binary Cyg OB2 \#8a and the magnetic star $\tau$\,Sco in the same reference. It should be noted, however, that $\theta^1$\,Ori\,C displays much higher temperatures (10--12MK for Mg and 15MK for Si). As \citet{wal07} noted, those temperatures agree well with the (ranges of) temperatures of maximum emissivities of these H- and He-like ions (this is also true for O, see below).

The $\cal G$ value that we derive for \hd\ indicates similar $\log(T)$ for the different elements: 6.95$\pm$0.25, 6.82$\pm$0.2 and 6.87$\pm$0.5 for S, Si and Mg, respectively. While the $\cal G$ ratios for Mg and Si are similar to those of OB stars (whatever their nature) reported by \citet{wal07,wal08}, the $\cal G$ ratio derived for S can only be compared to those of the magnetic objects $\tau$\,Sco and $\theta^1$\,Ori\,C, as well as to those of giant and main-sequence stars (only supergiants having higher ratios). It may be noted that the above temperatures corresponds to $\sim$0.7keV, which agrees well with the ``cool'' temperature of the ``hot'' global model (see next section). 

At temperatures $\log(T)$ of 6.8--7.0, ${\cal R}_0$ (calculated using ATOMDB as above) is about 2 for S, 2.65 for Si, and 2.42 for Mg. It is well known that ${\cal R} = \frac{f}{i}= {\cal R}_0\, \frac{1}{1 + \phi / \phi_c +  n_e / n_c}$, where we can neglect the density dependence for massive stars. The UV flux $\phi$ depends on the stellar output but also the dilution factor. We derive the UV flux for \hd\ using the model ($T_{eff}$=40kK and $\log(g)$=4.0) in the grid of O-star atmosphere models calculated with CMFGEN \citep{hil98}\footnote{http://kookaburra.phyast.pitt.edu/hillier/web/CMFGEN.htm} which is closest to the parameters derived from a dedicated atmosphere fitting of \hd\ \citep[and Table \ref{param}]{naz08}. After averaging the flux in the velocity interval [--2000,0]\kms near the rest wavelengths of the \nspec{2}{3}{S}{1} $\rightarrow$ \nspec{2}{3}{P}{1,2} transitions (following a similar path as  \citet{leu06}, see also Table\ref{wav}), we then derived a formation radius of 1.9$\pm$0.4\,$R_*$ from the $\cal R$ ratios of Si, and $<$3.2\,$R_*$ for Mg (the ratio itself corresponding to 1.7\,$R_*$)\footnote{Quoted errors take into account only the statistical errors of the measured fluxes of the f and i lines (and considering error propagation), not errors on the predicted CMFGEN EUV fluxes, which are difficult to quantify without actual observations of \hd\ at these EUV wavelengths.}. The sulphur lines are too noisy to provide a meaningful constraint on the formation radius. Such rather close radii are similar to those generally found for O-stars \citep{gue09}, including $\theta^1$\,Ori\,C \citep{gag05,gag05b}: therefore, it cannot be used to discriminate between various X-ray emission models. It may be worth noting, though, that our values fit well the temperature vs formation radius relation found by \citet{wal07,wal08}.

Following \citet{hue09}, we derive an abundance ratio Mg/Si, relative to the solar abundance of \citet{an89}, of 0.74$\pm$0.11 using the H-like and He-like resonance lines. Non-solar abundances in CNO elements are not entirely surprising since \hd\ displays both a nitrogen overabundance and an enriched surrounding nebula \citep[and references therein]{naz08}. However, changes in the Mg/Si ratio are not expected (see also next section).

Finally, we may try to use the ratio of H-to-He lines for oxygen to check whether our hypothesis of a cooler plasma is sensible. The O\,{\sc vii} triplet is very noisy in the RGS data, but the flux in the r line can be constrained to 2.4$\pm$0.6$\times10^{-5}$\,ph\,cm$^{-2}$\,s$^{-1}$ (observed) or $69\pm19\times10^{-5}$\,ph\,cm$^{-2}$\,s$^{-1}$ (corrected for absorption). This corresponds to a H-to-He ratio of 0.67$\pm$0.20, yielding a temperature of $\log(T)\sim6.35$. This is much lower than the temperatures found above: it would be difficult to reproduce this measurement in the context of an isothermal plasma, thereby confirming our hypothesis of a multi-temperature plasma. This result is certainly far from perfect because of its high uncertainty, but it constitutes a hint towards the non-uniform temperature of the X-ray emitting regions.

\section{Global fits}
Global fits were simultaneously made on the HEG, MEG and 0th order spectra, using a binning ensuring at least 20 counts per bin (see Sect. 2). Results of the fits are presented in Table \ref{tabglobal}. Note again that the quoted 1$\sigma$ errors are sometimes asymetrical: the value shown here always is the largest value. 

We used two sets of models that both assume thermal plasma in collisional ionization equlibrium (CIE). First, we used models with discrete temperature components ({\it apec+apec} or {\it apec+apec+apec} in Xspec). The second set models the emission of a distribution of radiative shocks as in \citet{zhp07}.
%We first tried to fit discrete temperature components ({\it apec+apec} or {\it apec+apec+apec} in Xspec), then we considered a distribution of optically-thin thermal plasma ({\it c6pvmkl} in Xspec, using the {\it apec} model interpolation), and XXXXXXXXXX. The second set models the emission of a distribution of radiative shocks as in \citet{zhp07}.

These emission models were multiplied by two absorption components: one corresponds to the fixed ISM absorption (model {\it wabs}, $N_{\rm H} = 4 \times 10^{21}$\,cm$^{-2}$, \citealt{dip94}), the other to potential additional local absorption (model {\it phabs}). Considering the results from the line fitting, the resulting absorption+emission spectrum was convolved with a constant-velocity Gaussian broadening using the {\it gsmooth} model in Xspec (width at 6\,keV=7.024\,eV and $\sigma(E)=\sigma(6keV)\times E/6$) while keeping the line shift to zero.

In each case, we first perform a fitting using solar abundances (with adopted abundances from \citealt{an89}) and we then varied the Ne, Mg, Si, S, and Fe abundances (i.e. those from elements displaying strong lines) to improve the fitting quality. Ne, Mg, and Si remain close to solar values, while S appears overabundant compared to solar and Fe subsolar. The major fit improvement in this case mostly comes from the changing iron abundance - the improvement from other elements is not formally significant. 

As found by \citet{naz08}, the presence of hot plasma is needed for a good representation of the X-ray emission from \hd. However, we found again the same ``duality'' in the 2T fits, with both a ``cool'' model (0.3\,keV+2\,keV) and a ``hot'' model (0.6\,keV+2.3\,keV) giving rather similar results. A three-temperature model rather yields best-fit plasma at 0.25, 0.8, and 2.4\, keV, but the improvement in $\chi^2$, compared to 2T models, is not very large. The relative importance of the two temperatures differs in these models, though: the 0.3\,keV component is largely dominant in the ``cool'' and 3T models, whereas the two temperatures of the ``hot'' model and the two higher temperatures of the 3T model have similar emission measures. As usual, there is clearly an interplay between temperatures, normalization factors and absorptions, but compared to ``normal'' O-stars \citep[e.g.][]{naz09}, the strength of the hottest component is much higher here. 

\begin{figure*}
\includegraphics[width=8cm]{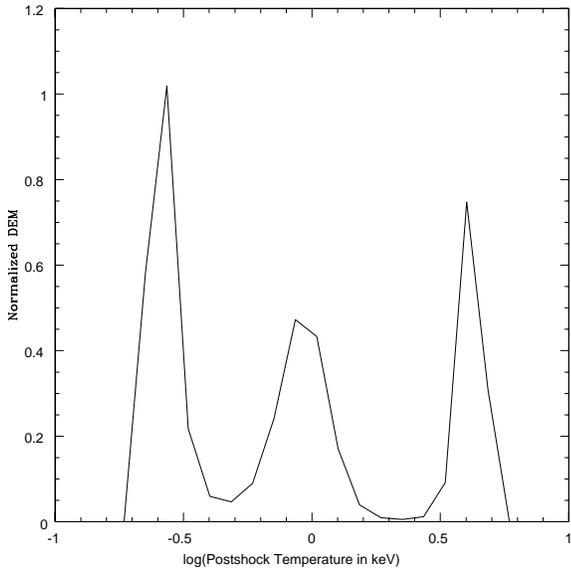}
\caption{The distribution of emission measure as a function of immediate postschock temperature for a model with distributed radiative shocks.
%The distribution of emission measure as a function of temperature for $c6pvmkal$ Xspec models (left, thin black solid line for solar abundances, thick red dashed line for non-solar abundances), XXX and distributed radiative shocks (right, thin black solid line for the former, thick red dashed line for the latter). 
\label{dem}}
\end{figure*}

The distributed shock model favor similar peak temperatures (0.25, 1. and 3\,keV, Fig. \ref{dem}), again in contrast to ``normal'' O-stars \citep{zhp07}, but with more contributions from the highest temperatures. This apparent difference in the shape of the distribution of emission measure of CIE plasma and that of radiative shocks deserves some comments. As described in \citet{zhp07}, at each temperature value the model with distribution of radiative shocks presents the total amount of emission measure of shocks with that given postshock temperature. Since the postshock region in a radiative shock is a temperature-stratified region by itself (occupied by plasma with temperature lower than the immediate postshock temperature, which is the one represented on Fig. \ref{dem}, explaining the skewing towards higher temperatures), then  the actual distribution of thermal plasma in the entire X-ray emitting region is different from that of the radiative shocks. 
%And this is what we derived in the case of \hd\ (Fig.~\ref{dem}). 
For such a reason, \citet{zhp07} emphasized that it is not correct to use {\it canonical} models of distribution of emission measure of CIE plasma to deduce the possible immediate postshock temperatures (and shock velocities) in the analysis of the X-ray emission from hot massive stars: the shocks  in their massive winds are expected to be radiative (see \S~2 in \citealt{zhp07} and the references therein).

Using the results from the global spectra fits, we can calculate the so-called total stellar `cloudiness' for \hd\ (see \S~5 and Appendix A in \citealt{zhp07}). Interestingly, we get a value of 0.008 which together with the zero-velocity line shift for its X-ray spectral lines (see previous section) puts \hd\ exactly next to $\theta^1$\,Ori\,C on the corresponding plot for a sample of hot massive stars (see Fig.~4 in \citealt{zhp07}). Indeed, when comparing the radiative shock model of \hd\ with the sample of \citet{zhp07}, it is obvious that the only object comparable to \hd\ is $\theta^1$\,Ori\,C (and maybe $\tau$\,Sco). However, whilst the temperature distributions of both objects for a radiative shock model show peaks at high temperatures ($kT>1$\,keV), \hd\ displays in addition a considerable amount of shock emission at temperatures below 1~keV - the emission below 1\,keV is rather negligible for $\theta^1$\,Ori\,C (compare Fig. \ref{dem} and Fig. 2 of \citealt{zhp07}). Similarly, the temperature distribution of CIE models show a smaller importance of the high-temperature peak compared to the low-temperature one for \hd\ (compare Table \ref{tabglobal} here and Table 2 in \citealt{gag05}). Indeed, a 2T fit of $\theta^1$\,Ori\,C yields values of 0.7 and 2.8\,keV with the hottest component having an emission measure 4 times larger than that of the coolest one, whereas the emission measures are similar for the ``hot'' model of \hd, which has temperatures similar to those of the 2T fit of $\theta^1$\,Ori\,C. This suggests that most of the overluminosity of \hd\ comes from relatively soft plasma.

In general, the absorption does not seem to exceed the ISM value by much. The derived abundance values are in general consistent with solar abundances, except for iron (but the solar abundance of iron has been revised since \citealt{an89}): the global fits confirm the 3/4 solar abundance of Mg/Si found from the line ratios. Finally, the flux observed by Chandra is consistent with that derived from the XMM-Newton low-resolution spectra \citep{naz08}, confirming the lack of large variations in the X-ray domain. 

\section{Summary and conclusion}
\hd\ is a peculiar object belonging to the intriguing category of Of?p stars. The presence of a magnetic field in these stars, together with their optical variations are generally explained within the framework of the Magnetic Oblique Rotator model. In 2008, a first X-ray investigation of \hd\ using XMM-Newton had revealed peculiarities in its X-ray spectrum too. Unfortunately, no detailed, high-resolution data of good quality was available until we obtained new data with Chandra-HETGS. 

These new data enable us to study \hd\ in more detail. The X-ray lines detected by Chandra display an average shift of $49\pm32$\,\kms, and an average FWHM of $827\pm90$\,\kms. The lines from high-Z elements, whose emissivity peaks at high temperatures, thus appear rather narrow, suggesting that they could be linked to magnetically confined winds. If the larger width of the O\,{\sc viii} line found from the noisier RGS data is real, it means that there is no unique source or location for the hot plasma. This is a similar situation to what was found for $\theta^1$\,Ori\,C \citep{gag05} - though the FWHM of the high-Z lines was even narrower for this star, about 600\,\kms. 

While RGS data already provided hints of a suppressed forbidden line for the Ne\,{\sc ix} triplet \citep{naz08}, the good quality HETG spectra confirm these hints and provide solid evidence for suppressed forbidden lines in two other He-like triplets, those of Si\,{\sc xiii} and Mg\,{\sc xii}. These He-like triplets of Mg and Si indicate a formation radius of about one stellar radius from the photosphere, which is rather close to the star's surface, but not uncommon in ``normal'' or magnetic O-stars \citep{gue09}. 

The temperatures derived from H-to-He like ratios are higher than for ``normal'' O-stars, but still not as extreme as those of $\theta^1$\,Ori\,C \citep{wal07,wal08}. This indicates the presence of hot plasma, which is confirmed in global fits.
% or, equivalently, to the peak temperature of the emission mesure in radiative shock models.
Global fits further show that this very hot component is much stronger than for ``normal'' O-stars but still much less dominant than in the extreme case of $\theta^1$\,Ori\,C \citep{zhp07,gag05}. The comparison of H-like and He-like line pairs supports this view, as \hd\ displays a higher ionization than ``normal'' O stars, albeit not quite as extreme as in $\theta^1$\,Ori\,C \citep{wal09}.

The fitting also shows that there is no large additional absorption over the ISM value, and that the overall flux agrees within 10\% with that recorded 8.5 years before by XMM-Newton. Regarding abundances of Ne, Mg, Si, and S, the global fits favor close-to-solar abundances (with a slight excess in sulphur and a deficit in iron). 

In conclusion, the new, good quality high-resolution data obtained by Chandra yielded additional clues in favor of the MOR model: relatively narrow X-ray lines (FWHM$\sim$800\kms) and a similar 'cloudiness' for the radiative shock model of \citet{zhp07}, plus a confirmation of the presence of very hot plasma. However, abundant soft plasma is also present, in contrast with the MOR-prototype $\theta^1$\,Ori\,C: in every respect (e.g. line widths of high-Z elements or temperature distributions), \hd\ thus appears less extreme than  $\theta^1$\,Ori\,C, despite their similar magnetic confinement parameters. Further MHD modelling of confined winds are definitely needed to explain this difference.

\acknowledgments
The authors thank Gregor Rauw and the referee for useful comments.
YN acknowledges support from the Fonds National de la Recherche Scientifique (Belgium), the PRODEX XMM and Integral contracts, and the `Action de Recherche Concert\'ee' (CFWB-Acad\'emie Wallonie Europe). YN also thanks W. Waldron, A. Foster, D. Huenemoerder and M. Leutenegger for helpful discussions. SAZ acknowledges financial support by NASA through Chandra grant GO0-11024A to the University of Colorado at Boulder, and publication costs were provided to NRW/STScI through Chandra grant GO0-11024B. The Space Telescope Science Institute is operated by AURA, Inc., under NASA contract NAS5-26555.

{\it Facilities:} \facility{CXO (HETG, ACIS)}, \facility{XMM (RGS)}.

\clearpage

\clearpage

\begin{table}
\begin{center}
\caption{Best-fit properties of the observed lines. \label{linefit}}
\begin{tabular}{lcccccc}
\tableline\tableline
Ion & $\lambda$ & E & shift  & FWHM  & Obs. Intensity & Cor. Intensity\\
 & (\AA) & (keV) & (\kms) & (\kms) &  (ph\,cm$^{-2}$\,s$^{-1}$) &  (ph\,cm$^{-2}$\,s$^{-1}$) \\
\tableline
Chandra\\
S\,{\sc xv}   & 5.03873 &2.46062 &   88$\pm$225 & 962$\pm$809 & (9.3$\pm$2.8)$\times 10^{-6}$   &(10.2$\pm$3.1)$\times 10^{-6}$   \\
            & 5.06649 &2.44714 &              &             & (2.6$\pm$2.6)$\times 10^{-6}$   & (2.9$\pm$2.9)$\times 10^{-6}$   \\
            & 5.10150 &2.43035 &              &             & (5.5$\pm$2.6)$\times 10^{-6}$   & (6.1$\pm$2.9)$\times 10^{-6}$   \\
Si\,{\sc xiv} & 6.18044 &2.00607 & 0$\pm$64     & 805$\pm$252 & (9.30$\pm$1.08)$\times 10^{-6}$ &(11.0$\pm$1.28)$\times 10^{-6}$ \\
            & 6.18585 &2.00432 &              &             & \multicolumn{2}{c}{previous$\times$0.509}                       \\
Si\,{\sc xiii}& 6.64795 &1.86500 &   117$\pm$59 & 859$\pm$160 & (1.46$\pm$0.18)$\times 10^{-5}$ & (1.80$\pm$0.22)$\times 10^{-5}$ \\
            & 6.68819 &1.85378 &              &             & (7.52$\pm$1.43)$\times 10^{-6}$ & (9.30$\pm$1.77)$\times 10^{-6}$ \\
            & 6.74029 &1.83945 &              &             & (6.02$\pm$1.30)$\times 10^{-6}$ & (7.32$\pm$1.58)$\times 10^{-6}$ \\
Mg\,{\sc xii} & 8.41921 &1.47263 &   33$\pm$57  & 769$\pm$152 & (8.78$\pm$1.15)$\times 10^{-6}$ & (12.6$\pm$1.65)$\times 10^{-6}$ \\
            & 8.42462 &1.47169 &              &             & \multicolumn{2}{c}{previous$\times$0.507}\\
Mg\,{\sc xi}  & 9.16875 &1.35225 &   42$\pm$123 & 894$\pm$297 & (1.01$\pm$0.25)$\times 10^{-5}$ & (1.59$\pm$0.39)$\times 10^{-5}$ \\
            & 9.23121 &1.34310 &              &             & (7.0$\pm$2.6)$\times 10^{-6}$   &(11.1$\pm$4.1)$\times 10^{-6}$   \\
            & 9.31434 &1.33111 &              &             & (0.5$\pm$1.4)$\times 10^{-6}$   & (0.8$\pm$2.3)$\times 10^{-6}$   \\
Ne\,{\sc x}   &12.13210 &1.02195 &--15$\pm$150  & 876$\pm$275 & (1.73$\pm$0.33)$\times 10^{-5}$ & (4.31$\pm$0.82)$\times 10^{-5}$ \\
            &12.13750 &1.02150 &              &             & \multicolumn{2}{c}{previous$\times$0.502}\\
\tableline
XMM \\
Ne\,{\sc x}   &12.13210 &1.02195 &--18$\pm$510  &1815$\pm$1124& (4.0$\pm$0.7)$\times 10^{-5}$ &(10.0$\pm$1.7)$\times 10^{-5}$ \\
            &12.13750 &1.02150 &              &             & \multicolumn{2}{c}{previous$\times$0.502}\\
O\,{\sc viii} &18.96710 &0.65368 &--440$\pm$180 &1537$\pm$505 & (2.9$\pm$0.4)$\times 10^{-5}$ &(30.7$\pm$4.2)$\times 10^{-5}$ \\
            &18.97250 &0.65349 &              &             & \multicolumn{2}{c}{previous$\times$0.501} \\
\tableline
\end{tabular}
\end{center}
\end{table}

\begin{table}
\begin{center}
\caption{Stellar parameters (from \citealt{naz08}). Note that the wind parameters remain uncertain. \label{param}}
\begin{tabular}{lc}
\tableline\tableline
Parameter & Value \\
\tableline
$T_{eff}$ & 41$\pm$2\,kK \\
$\log g$ & 4.0$\pm$0.1 \\
$v_{\infty}$ & $\sim$2600\kms\\
$\log (\dot M / \sqrt{f})$ & $-6$\\
\tableline
\end{tabular}
\end{center}
\end{table}

\begin{table}
\begin{center}
\caption{Observed line ratios. \label{linerat}}
\begin{tabular}{lcccc}
\tableline\tableline
Element & R & G & H-to-He & H-to-He (r only)\\
\tableline
S & 2.1$\pm$2.3 & 0.88$\pm$0.48 & &\\
Si& 0.79$\pm$0.23 & 0.92$\pm$0.17 & 0.48$\pm$0.07 & 0.92$\pm$0.16 \\
Mg& 0.07$\pm$0.21 & 0.75$\pm$0.35 & 0.68$\pm$0.17 & 1.19$\pm$0.33 \\
\tableline
\end{tabular}
\end{center}
\end{table}

\clearpage
\begin{table}
\begin{center}
\caption{Rest wavelengths of the \nspec{2}{3}{S}{1} \(\rightarrow\) \nspec{2}{3}{P}{1,2} transitions in ATOMDB v 2.0.0. \label{wav}}
\begin{tabular}{lcc}
\tableline\tableline
Ion & $\lambda_1$(\AA, \nspec{2}{3}{S}{1} \(\rightarrow\) \nspec{2}{3}{P}{1}) & $\lambda_2$(\AA, \nspec{2}{3}{S}{1} \(\rightarrow\) \nspec{2}{3}{P}{2})\\
\tableline
S\,{\sc xv}    & 738.3 & 673.4 \\
Si\,{\sc xiii} & 865.1 & 814.7 \\
Mg\,{\sc xi}   & 1034.3& 997.5 \\
\tableline
\end{tabular}
\end{center}
\end{table}

%\clearpage

\begin{table}
\begin{center}
\caption{Results of the fitting. Abundances are in numbers, relative to hydrogen and relative to solar. \label{tabglobal}}
\begin{tabular}{lccccccc}
\tableline\tableline
Parameter & \multicolumn{2}{c}{2T$_{\rm cool}$} & \multicolumn{2}{c}{2T$_{\rm hot}$} & \multicolumn{2}{c}{3T} & Rad. shocks\\
& (solar) & (var.) &  (solar) & (var.) &  (solar) & (var.) &  \\
\tableline
$N_{\rm H}$ (10$^{22}$\,cm$^{-2}$)         & 0.34$\pm$0.03 & 0.14$\pm$0.04 & 0.13$\pm$0.02 &   0.$\pm$0.01 & 0.27$\pm$0.03 & 0.13$\pm$0.06 & 0.11$\pm$0.07 \\
$kT_1$ (keV)                               & 0.28$\pm$0.01 & 0.37$\pm$0.02 & 0.64$\pm$0.01 & 0.65$\pm$0.01 & 0.24$\pm$0.02 & 0.30$\pm$0.03 &           \\
$norm_1$ (10$^{-3}$\,cm$^{-5}$)            & 10.9$\pm$1.47 & 4.47$\pm$0.90 & 1.89$\pm$0.16 & 1.78$\pm$0.16 & 7.28$\pm$2.30 & 2.89$\pm$1.25 &  8.25             \\
$kT_2$ (keV)                               & 1.91$\pm$0.04 & 2.04$\pm$0.06 & 2.34$\pm$0.06 & 2.44$\pm$0.08 & 0.78$\pm$0.03 & 0.79$\pm$0.04 &               \\
$norm_2$ (10$^{-3}$\,cm$^{-5}$)            & 2.66$\pm$0.06 & 2.39$\pm$0.09 & 1.93$\pm$0.05 & 1.83$\pm$0.07 & 1.53$\pm$0.15 & 1.31$\pm$0.19 &               \\
$kT_3$ (keV)                               &               &               &               &               & 2.33$\pm$0.08 & 2.52$\pm$0.10 &               \\
$norm_3$ (10$^{-3}$\,cm$^{-5}$)            &               &               &               &               & 1.91$\pm$0.07 & 1.72$\pm$0.08 &               \\
Ne                                         &               & 0.92$\pm$0.12 &               & 0.96$\pm$0.14 &               & 0.90$\pm$0.15 & 0.86$\pm$0.16 \\
Mg                                         &               & 0.84$\pm$0.15 &               & 0.86$\pm$0.10 &               & 0.99$\pm$0.12 & 0.85$\pm$0.12 \\
Si                                         &               & 1.15$\pm$0.11 &               & 1.07$\pm$0.11 &               & 1.15$\pm$0.12 & 1.03$\pm$0.11 \\
S                                          &               & 1.41$\pm$0.19 &               & 1.73$\pm$0.22 &               & 1.82$\pm$0.24 & 1.86$\pm$0.25 \\
Fe                                         &               & 0.53$\pm$0.06 &               & 0.59$\pm$0.06 &               & 0.75$\pm$0.09 & 0.71$\pm$0.11 \\
Flux (10$^{-12}$\,erg\,cm$^{-2}$\,s$^{-1}$)& 2.70          & 2.70          & 2.73          & 2.76          & 2.77          & 2.79          & 2.99          \\
$\chi^2$ (dof)                             & 1.02 (556)    & 0.94 (551)    & 0.84 (556)    & 0.74 (551)    & 0.73 (554)    & 0.69 (549)    & 0.61 (534)    \\
\tableline
\end{tabular}
\end{center}
\end{table}

\end{document}